\documentclass[prb,preprint,amssymb,showpacs]{revtex4}
\usepackage{bm}
\usepackage{graphicx}
\usepackage{dcolumn}

\begin{document}

\title{Electronic friction for a slow molecule in a metal}

\date{\today}

\author{A. Salin\footnote{Retired from Universit\'e de Bordeaux I}}
\affiliation{12, rue Jules Testaud, 33700 M\'erignac, France}
\email{asalin@mailaps.org}

\begin{abstract}
In a previous contribution, we have set up the framework for the calculation of electronic friction for a slow atom traveling through a metal. We provide in the present work a generalization to the case of polyatomic molecules. This formulation is restricted to those molecular degrees of freedom for which a semi-classical treatment of friction is valid.
\end{abstract}
\maketitle

%==============================================================================
\section{Introduction}

In a previous contribution\cite{previous} we have set up a procedure for the determination of the electronic friction experienced by slow atoms moving through a metal. Electronic friction originates from the fact that, in a metal, electronic excitations by a moving particle may take place with a vanishingly small energy change. This gives rise, at low projectile velocities, to a linear dependence of the electronic stopping power on the velocity, $\bm{v}$, of the atom. The coefficient of this linear behavior is called the {\em friction coefficient}. By definition, the determination of the friction coefficient corresponds to a first order approximation in the projectile velocity. This has allowed us to prove that it can be obtained exactly (in principle) from a static ensemble Kohn-Sham procedure, the ensemble being defined by shifting the Fermi surface by $-\bm{v}$.

In the present work, we extend this previous procedure to the case of molecules. Indeed, electronic friction may play an important role in various molecular processes, on or in metals: molecular adsorption, vibrationnal or rotationnal relaxation, etc. Until now, the only reliable calculations of molecular friction have made use of the Independent Atom approximation\cite{jua08,mar12}. Whereas it is reasonable to believe that the latter approximation gives the correct order of magnitude, it is highly desirable to gauge its validity. We set up in the present work the framework that allows such calculations.

Atomic units are used throughout.
%==============================================================================
\section{Theory}

As in our previous work \cite{previous}, the medium in which the molecule moves is composed of electrons in a periodic array of fixed
nuclei. We are interested in the projectile energy loss due to the electronic
state perturbation by the projectile motion. The interaction of the molecule with the nuclei of the
frozen lattice is not considered in the following development. Our
task is therefore to determine the electronic state for the electrons moving in the field of the fixed lattice nuclei and the moving ones of the molecule. Again we consider the low energy regime, i.e. vanishingly small velocities (with respect to the lattice) for all nuclei in the molecule.

If all internal ro-vibrationnal states of the molecule are frozen, then the theory developed for an atom can be applied. The transformation from the lattice to the molecular frame can be performed by shifting the Fermi surface by $-\bm{v}$, where $\bm v$ is the velocity of the molecule. Then, to first order in the velocity, the density of the system can be obtained by a static ensemble Kohn-Sham procedure which yields to an exact (in principle) determination of the friction coefficient. 

If the internal degrees of freedom are not frozen, then the formulation of the relaxation of internal motion in terms of a friction coefficient is not strictly valid because of the discrete spectrum associated with them. Indeed, the transfer of energy from internal degrees of freedom to the electrons is not a continuous process as implicitly assumed when using the Shifted Fermi Surface procedure. As a consequence, an associated friction coefficient only makes sense if the considered internal motion can be treated as quasi-continuous and a semi-classical treatment is appropriate. If not, the existence of an excitation threshold precludes a linear dependence on velocity: obviously, such a behavior can only be associated with a continuous spectrum of allowed energy transfers.

The energy loss associated with one degree of freedom of the molecule is in principle dependent on the motion, i.e., velocities, associated with all other degrees of freedom. For example, in the case of a diatomic molecule AB, the friction associated with the motion of atom A along the direction ${\hat{\bm v}}_A$ depends on both $\bm{v}_A$ {\em and} $\bm{v}_B$, so that it is a 6$\times$6 tensor. More generally, if the molecule consists in $N$ nuclei, the energy-loss is determined by a $3N\times 3N$ tensor. The determination of such a quantity looks like a formidable problem.

To simplify our development, we begin with the case of a diatomic molecule and consider subsequently the generalization to polyatomic molecules.
 
%++++++++++++++++++++++++++++++++++++++++++++++++++++++++++++++++++++++++++++++
\subsection{Diatomic molecule}

The position vector of the nuclei are $\bm{R}_1$ and $\bm{R}_2$ respectively. The  nucleus 1 (resp.\ 2) moves with velocity $\bm{v}_1$ (resp.\ $\bm{v}_2$). The electronic energy of the system is $E(\bm{R}_1,\bm{R}_2;\bm{v}_1,\bm{v}_2)$. In the present semi-classical context, we denote as $\bm{F}_1$ (resp.\ $\bm{F}_2$) the force on nucleus 1 (resp.~2):
\begin{equation}
\bm{F}_{1,2}(\bm{R}_1,\bm{R}_2;\bm{v}_1,\bm{v}_2) =  - \bm{\nabla}_{\bm{R}_{1,2}}\,E(\bm{R}_1,\bm{R}_2;\bm{v}_1,\bm{v}_2) 
\label{eq1} \end{equation}
As the electronic state is not homogeneous, the nuclei experience a force (adiabatic force) even when they are at rest with respect to the crystal lattice.
Let us denote the electronic density as $n_{\bm{v}_1,\bm{v}_2}(\bm{r};\bm{R}_1,\bm{R}_2)$. Using Hellmann-Feynmann theorem:
\begin{equation}
\bm{F}_{1,2}(\bm{R}_1,\bm{R}_2;0,0) =  -\int d\bm{r}\, n_{0,0}(\bm{r};\bm{R}_1,\bm{R}_2)\, \bm{\nabla}_{\bm{R}_{1,2}} V_{eP_{1,2}}(\bm{r};\bm{R}_{1,2})  
\label{eq2} \end{equation}
where $V_{eP_1}$ (resp.\ $V_{eP_2}$) is the interaction potential between the electrons and nucleus 1 (resp.~2).

Consider first that both nuclei move with the same velocity $\bm{v}$. Then, the theory developed for an atom in Ref.~\onlinecite{previous} can be applied, as noted above: to first order in the velocity, the density can be evaluated through a static ensemble Kohn-Sham procedure and the energy functional is stationary with respect to the variation of any parameter of the system. As a consequence, the force on either nucleus can be expressed as:
\begin{equation}
\bm{F}_{1,2}(\bm{R}_1,\bm{R}_2;\bm{v},\bm{v}) =  -\int d\bm{r}\, n_{\bm{v},\bm{v}}(\bm{r};\bm{R}_1,\bm{R}_2)\, \bm{\nabla}_{\bm{R}_{1,2}} V_{eP_{1,2}}(\bm{r};\bm{R}_{1,2})  
\label{eq3}\end{equation}  
and each of them contributes to the friction through a friction coefficient along the direction~$\hat{\bm{v}}$:
\begin{equation}
{\cal{F}}_{\hat{\bm{v}}}^{1,2} = \lim_{v\rightarrow 0}{1\over v}\, \int d\bm{r}\,\left[ n_{\bm{v},\bm{v}}(\bm{r};\bm{R}_1,\bm{R}_2)-n_{0,0}(\bm{r};\bm{R}_1,\bm{R}_2)\right]\,\hat{\bm{v}}\cdot\bm{\nabla}_{\bm{R}_{1,2}}V_{eP_{1,2}}(\bm{r};\bm{R}_{1,2})
\label{eq4}\end{equation}
Test calculations using (\ref{eq4}) for H$_2$ and HLi traveling parallel to their internuclear axis through an homogeneous electron gas (jellium) give results in agreement with those of Ref.~\onlinecite{die00}. 

It is not obvious that (\ref{eq4}) is valid when the two nuclei are moving with different velocities. We consider now the latter case.
From (\ref{eq2}) the friction coefficient for nucleus 1 and 2, along the direction of their velocity, is given by:
\begin{eqnarray}
{\cal{F}}_{\hat{\bm{v}}_1}^1 &=& \lim_{v_1\rightarrow 0}{1\over v_1}\, \hat{\bm{v}}_1\cdot\bm{\nabla}_{\bm{R}_1}\left[E(\bm{R}_1,\bm{R}_2;\bm{v}_1,\bm{v}_2) - E(\bm{R}_1,\bm{R}_2;0,\bm{v}_2) \right]\\
{\cal{F}}_{\hat{\bm{v}}_2}^2 &=& \lim_{v_2\rightarrow 0}{1\over v_2}\, \hat{\bm{v}}_2\cdot\bm{\nabla}_{\bm{R}_2}\left[E(\bm{R}_1,\bm{R}_2;\bm{v}_1,\bm{v}_2) - E(\bm{R}_1,\bm{R}_2;\bm{v}_1,0)  \right]
\label{eq6} \end{eqnarray}
We wish to determine the latter quantities to zero order in $v_1$ and $v_2$. We note that:
\begin{equation}
E(\bm{R}_1,\bm{R}_2;\bm{v}_1,\bm{v}_2) - E(\bm{R}_1,\bm{R}_2;0,\bm{v}_2) = \left.\bm{v}_1\cdot\bm{\nabla}_{\bm{v}_1} E(\bm{R}_1,\bm{R}_2;\bm{v}_1,\bm{v}_2)\right|_{v_1=0}+ 0(v_1^2)
\label{eq7} \end{equation}
In addition:
\begin{eqnarray}
\hspace{-1.cm}\left.\bm{v}_1\cdot\bm{\nabla}_{\bm{v}_1} E(\bm{R}_1,\bm{R}_2;\bm{v}_1,\bm{v}_2)\right|_{v_1=0}
 &=& \left.\bm{v}_1\cdot\bm{\nabla}_{\bm{v}_1} E(\bm{R}_1,\bm{R}_2;\bm{v}_1,0)\right|_{v_1=0} \hspace*{3cm}\nonumber \\
& &  \hspace{-.2cm} + \left. \left[\bm{v}_1\cdot\bm{\nabla}_{\bm{v}_1}\right] \left[ \bm{v}_2\cdot\bm{\nabla}_{\bm{v}_2}\right] 
E(\bm{R}_1,\bm{R}_2;\bm{v}_1,\bm{v}_2)\right|_{v_1,v_2=0} + 0(v_2^2)\\
&=& \left.\bm{v}_1\cdot\bm{\nabla}_{\bm{v}_1}  E(\bm{R}_1,\bm{R}_2;\bm{v}_1,0)\right|_{v_1=0} + 0(v_2) \\
&=& E(\bm{R}_1,\bm{R}_2;\bm{v}_1,0) - E(\bm{R}_1,\bm{R}_2;0,0) + 0(v_1,v_2)
\label{eq8} \end{eqnarray}
Finally:
\begin{equation}
{\cal{F}}_{\hat{\bm{v}}_1}^1 = \lim_{v_1\rightarrow 0}{1\over v_1}\,\hat{\bm{v}}_1\cdot \bm{\nabla}_{\bm{R}_1}\left[E(\bm{R}_1,\bm{R}_2;\bm{v}_1,0) - E(\bm{R}_1,\bm{R}_2;0,0) \right] + 0(v_1,v_2)
\label{eq9}\end{equation}
The latter expression is that of the friction for nucleus 1 moving alone while the other nucleus is at rest with respect to the lattice, therefore playing the same role as the lattice nuclei. Hence, we may apply the theory developed in Ref.~\onlinecite{previous} for a single moving nucleus and the friction can be expressed as:
\begin{equation}
 {\cal{F}}_{\hat{\bm{v}}_1}^1= \lim_{v_1\rightarrow 0}{1\over v_1} \int d\bm{r}\, \left[n_{\bm{v}_1,0}(\bm{r};\bm{R}_1,\bm{R}_2)-n_{0,0}(\bm{r};\bm{R}_1,\bm{R}_2)\right]\,  \hat{\bm{v}}_1\cdot\bm{\nabla}_{\bm{R}_1}  V_{eP_1}(\bm{r};R_1)  
\label{eq10} \end{equation}
A similar derivation can be carried out for nucleus 2.

This results allows to draw various conclusions. Firstly, the friction coefficient involves only ``local'' quantities (beside the derivative of $V_{eP}$), i.e., it does not require information on the {\em variation} of any quantity characterizing the system electronic state. 

Secondly, the friction is completely determined by 6 independent quantities associated with three orthogonal velocity directions for each nucleus. When using a different coordinate system (e.g., cartesian coordinates for the center of mass motion and spherical coordinates for the molecular internal motion) the friction involves a $6\times 6$ tensor, but the latter is completely determined by the previous 6 independent values. The expression of the $6\times 6$ tensor in terms of the 6 cartesian components is formally the same as for the Independent Atom approximation in which the molecule is assumed to be composed of two independent atoms \cite{inaki}. Of course, the 6 friction components calculated from (\ref{eq10}) have not the same value as in the Independent Atom approximation since the density entering (\ref{eq10}) involves both nuclei within the lattice. It should be kept in mind that, although the previous procedure allows to calculate a friction coefficient associated with any degree of freedom, the latter only makes sense if the dynamics of this degree of freedom may be described by a semi-classical approximation.

Thirdly, in (\ref{eq10}), the velocity enters only through the shift in the Fermi Surface. Therefore the density is the same for nucleus 1 moving with velocity $\bm{v}$ and for nucleus 2 moving with velocity $\bm{v}$. Accordingly, only one calculation of the density is required for a given velocity orientation. In practice, calculations for more than one value of $\bm v$ are required to check the linear behavior of the force with respect to velocity. However, this is independent of the fact that we consider a molecule rather than an atom.

%++++++++++++++++++++++++++++++++++++++++++++++++++++++++++++++++++++++++++++++
\subsection{Polyatomic molecule}

The extension to a polyatomic molecule ($N$ nuclei) is trivial and expression (\ref{eq10}) can be extended to evaluate the friction for any nucleus $i$:
\begin{equation}
 {\cal{F}}_{\hat{\bm{v}}}^i= \lim_{v\rightarrow 0}{1\over v} \int d\bm{r}\, \left[n_{\bm{v}}(\bm{r};\bm{R}_1,\ldots,\bm{R}_N)-n_0(\bm{r};\bm{R}_1,\ldots,\bm{R}_N)\right]\,  \hat{\bm{v}}\cdot\bm{\nabla}_{\bm{R}_i}  V_{eP_i}(\bm{r};R_i)  
\label{eq11} \end{equation}
Here the density is noted as $n_{\bm{v}}$ since the velocity only appears through the shift in the Fermi surface and is not necessarily associated with a given atom, provided we are in the linear regime for the force. The friction for all nuclei is entirely determined by 3$N$ quantities associated with three orthogonal velocity directions for each nucleus. When using any alternative coordinate system, the $3N\times3N$ tensor can be expressed in terms of these $3N$ quantities. 

%==============================================================================
\section{Conclusion}

Our derivation proves that the main conclusion of our work on atoms can be extended to the case of polyatomic molecules. Electronic friction can be determined by a static ensemble Kohn-Sham procedure using a shifted Fermi surface to set the ensemble of occupied Kohn-Sham orbitals. This is a direct consequence of the fact that the density can be determined exactly under this procedure in a first order approximation with respect to nuclear velocities. In addition, we have shown that, whereas friction is, in general, expressed in terms of a $3N\times 3N$ tensor ($N$ being the number of nuclei in the molecule), only $3N$ quantities are needed to determine completely this tensor. Finally a single calculation of the system electronic density is required to determine the complete set of $3N$ independent coefficients. Thus, what seemed a formidable and, even, untractable problem is amenable to a solution. Of course, what remains to be achieved is the implementation of the shifted Fermi surface procedure within the band structure codes.


\begin{thebibliography}{99}

\bibitem{previous} A.~Salin, arXiv:1302.0986 [cond-mat.other].
\bibitem{jua08} J.I.~Juaristi, M.~Alducin, R.~D\'\i ez-Mui\~ no, H.F.~Busnengo, and A.~Salin, Phys.\ Rev.\ Lett.\ {\bf 100}, 116102 (2008).
\bibitem{mar12} L.~Martin-Gondre, M.~Alducin, G.A.~Bocan, R.~D\'\i ez~Mui\~no, and J.~I.~Juaristi, Phys.\ Rev.\ Lett.\ {\bf 108}, 096101 (2012).
\bibitem{die00} R.~D\'\i ez~Mui\~no and A.~Salin, Phys.\ Rev.\ B {\bf 62}, 5207 (2000).
\bibitem{inaki}  J.I.~Juaristi, private communication.
\end{thebibliography}
\end{document}